# Improving the temporal resolution of event-based electron detectors using neural network cluster analysis

Alexander Schröder[1,2], Leon van Velzen[3], Maurits Kelder[3], Sascha Schäfer[1,2, *]

[1] Institute of Physics, University of Oldenburg, Oldenburg, Germany

[2] Department of Physics, University of Regensburg, Regensburg, Germany

[3] Amsterdam Scientific Instruments (ASI), Amsterdam, Netherlands

* Corresponding author: Sascha.schaefer@physik.uni-regensburg.de

Novel event-based electron detector platforms provide an avenue to extend the temporal resolution of electron microscopy into the ultrafast domain. Here, we characterize the timing accuracy of a detector based on a TimePix3 architecture using femtosecond electron pulse trains as a reference. With a large dataset of event clusters triggered by individual incident electrons, a neural network is trained to predict the electron arrival time. Corrected timings of event clusters show a temporal resolution of 2 ns, a 1.6-fold improvement over cluster-averaged timings. This method is applicable to other fast electron detectors down to sub-nanosecond temporal resolutions, offering a promising solution to enhance the precision of electron timing for various electron microscopy applications.

## I. Introduction

In the recent decade, ultrafast transmission electron microscopy (UTEM) based on pico- and femtosecond electron pulses has made a tremendous progress[1-10], reaching down to an attosecond temporal resolution[11,12] and opening new avenues for quantum electron microscopy[13-15]. In a complementary approach, fast electron cameras are being developed[16] which would allow pico- to nanosecond-scale imaging with continuous electron beams at higher average beam currents compared to UTEM. For example, a delay-line detector combined with microchannel plate (MCP) amplification demonstrated a temporal resolution of 122 ps, enabling the imaging of the gyrational motion of a magnetic vortex[17]. A further detector platform involves large arrays of time-to-digital converters (TDC) as implemented in the TimePix3 chip architecture[18] and originally employed for X-ray detection[19,20]. First applications of the TimePix3 detector in electron imaging involves the energy- and angle-resolved detection of low-energy photoelectrons[21], as well as protein imaging[22,23] and coincidence measurements[24] at high electron energies. The intrinsic temporal binning width of the TimePix3 chip allows for a temporal resolution of 1.56 ns.
By utilizing the intrinsic correlations between the triggering time and the duration of the above-threshold signal, a root-mean-square (rms) resolution of 1.7 ns, close to the binning width, was demonstrated for low-energy electrons[25]. At higher electron energies, the substantially different electron-sensor interactions are expected to generate a more complex temporal detector response, which is not, yet, experimentally characterized.

Here, we utilize femtosecond electron pulses in an ultrafast transmission electron microscope to determine the temporal response and event correlation properties of a TimePix3 electron detector at 200 keV electron energies. A neural network approach is presented to improve the temporal resolution of electron event detection. The neural network is trained based on experimental data recorded for femtosecond electron pulses. Corrected timings of event clusters show a temporal width of 2 ns (rms), increasing the achievable time resolution by a factor of 1.6 compared to cluster-averaged timings.

## II. Experimental Approach

The event-based electron detector used in the experiments (CheeTah T3, Amsterdam Scientific Instruments) utilizes the TimePix3 architecture[18] and is composed of four individual chips (512 x 512 pixels, 28 mm x 28 mm sensor size) bump-bonded to a 300-µm thick silicon sensor (Fig. 1a). A high-energy electron incident on the sensor locally generates free carriers and secondary electrons resulting in a signal peak in the input channels of close-by pixels in the TimePix chip. Within each pixel, an event is registered if the incoming signal exceeds a threshold value. A time stamp (time-of-arrival, ToA) is assigned to such an event using a global and local clock operating at frequency of 40 MHz and 640 MHz, respectively. In addition, the time-over-threshold (ToT) for each event is determined. ToA, ToT and the pixel coordinates of an event are buffered and finally stored on a hard drive. Each double-pixel column has a counter that is reset at the start of acquisition. Due to a known reset synchronization issue, about 5 percent of pixel columns are shifted in their time response by one period of the coarse clock (25 ns), which we manually correct.

For assigning a set of registered events to a single incident electron, a cluster algorithm is used, which groups events occurring in a defined time interval (5 µs) within an 8x8 pixel area.

The TimePix3 camera is incorporated into the Oldenburg ultrafast transmission electron microscope (Ol-UTEM), which is based on a JEOL F200 Schottky field-emitter TEM. A modified laser-driven electron source[7] allows for the generation of femtosecond electron pulses (200-keV electron energy, 200-fs pulse duration, 400-kHz repetition rate) using ultrashort ultraviolet light pulses (200-fs temporal width, 343 nm center wavelength). The electron pulses arriving at the camera are spread across the whole detector area with an approximately homogeneous intensity. On average, 1.3 event clusters are detected on the camera per optical pulse.

An electronic trigger signal synchronized to the electron pulse train is generated by a fast photodiode (12 GHz bandwidth) illuminated by the photoemission laser and fed into the TimePix3 event stream. The electron pulse duration and any jitter in the synchronous trigger signal are significantly shorter than the 1.56-ns clock period employed in the TimePix3 architecture. Therefore, the temporal spread of registered electron events directly signifies the temporal resolution of the detector.

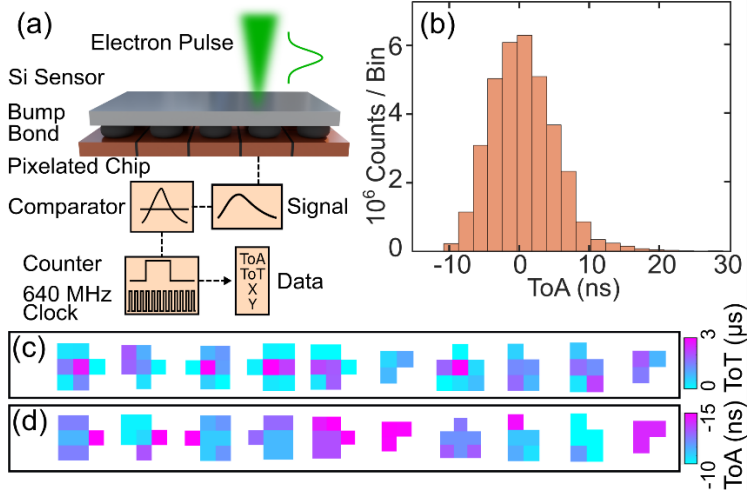

**Figure 1: Measurement principle and event timings.** (a) Experimental setup and schematic of TimePix3 event detection. (b) Uncorrected time-of-arrival distribution of photoelectron pulses relative to an optical trigger signal (averaged over 4 x $10^6$ electron pulses) showing temporal of about 10 ns (FWHM). (c) Examples of randomly selected electron event clusters each induced by a single incident electron; color scales denote time-over-threshold (top) and time-of-arrival (bottom).

## III. Electron Detection

For characterizing the temporal response of the TimePix3 electron detector, we first consider the histogram of event timings relative to the photodiode trigger signal, as displayed in Fig. 1(b) with the mean ToA of the distribution set to zero. The temporal response function exhibits a full-width-at-half-maximum (FWHM) of approximately 10 ns (4.2 ns, rms), considerably longer than the electron pulse width and the temporal bin width of the TimePix detector of 1.56 ns. In addition, the distribution function is skewed towards larger delays.

The temporal accuracy can be improved by considering event clusters triggered by a single incident electron instead of the individual events in each pixel. Typical, randomly selected event clusters are shown in Fig. 1 (c,d). For the employed detector settings (about 15-keV projected threshold, 100-V Bias), each incident electron triggers a cluster of 6.5 events on average. The cluster shape is strongly varying putatively due to the random scattering processes of the incident electron within the silicon sensor. Furthermore, within a cluster, time-of-arrival and time-over-threshold values differ by large amounts, which can be partially traced back to a time-walk effect, given the fixed threshold level and the varying signal amplitudes arriving at each pixel. However, different from previous results for the combination of a TimePix3 chip with a light-sensitive sensor[21,25], we find no clear correlation between the time-of-arrival and time-over-threshold values registered in a pixel. This is further evidenced by considering the joint probability distribution between both quantities as shown in Fig. 2(a). Instead, we observe distinct features in joint distributions connecting the number of involved pixels in an event cluster, i.e. the cluster size, and the summed ToT and averaged ToA within the cluster, respectively. As shown in Fig. 2(b), the distribution of cluster-summed ToT values consists of two components. One intense structure which shows narrow a ToT distribution independent of cluster size, and a structure corresponding to a smaller number of clusters for which the summed ToT increased approximately linearly with cluster size.

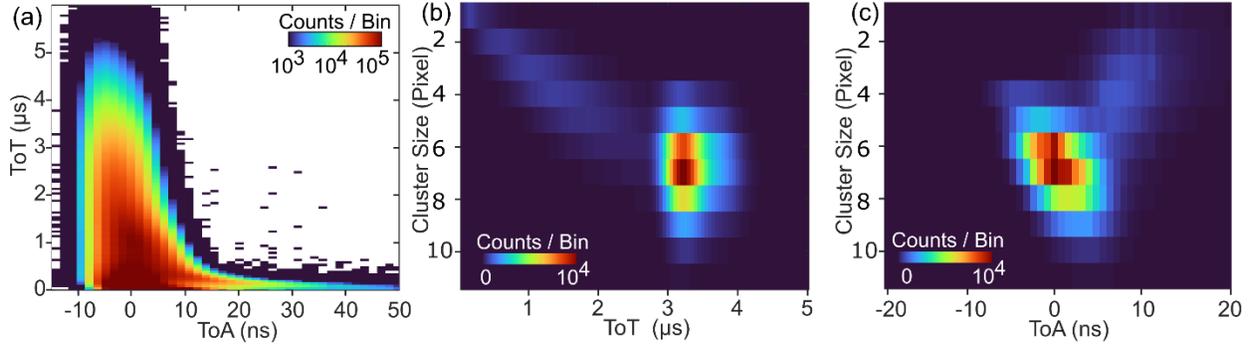

**Figure 2: Correlations between event cluster properties.** (a) Two-dimensional histogram of time-of-arrival and time-over-threshold evaluated for individual events. (b-c) Correlation between the number of above-threshold pixels in an event cluster size and the summed time-over-threshold (b) and averaged time-of-arrival (c).

Similarly, also for the joint distribution of ToAs and cluster sizes both features are found (Fig. 2c). Here, the dominant distribution component shifts in its maximum with increasing cluster size in line with an expected time-walk effect. A second, less intense component is observed for small cluster sizes with a slight tilt in the opposite direction as compared to the first component. A likely cause for these features are signals stemming from secondary electrons created within the sensor material and slowed-down primary electrons hitting the TimePix3 detector electronics. Summing the joint distributions along the direction of the cluster size as a random variable and neglecting the intrinsic correlations results in the broadened overall ToA distribution, as displayed in Fig. 1(b), and a concomitant loss of temporal resolution. A more precise timing information is achieved by compensating for the linear shift in the main feature of Fig. 2(c) with cluster size. The resulting corrected ToA distribution is plotted in Fig. 3(b) (blue curve).

## IV. Improving temporal resolution

In an ideal case, all of the intrinsically contained correlations within the event data set should be exploited to extract a most faithful estimate of the arrival time of the incident primary electron. To approximate this goal, we trained a neural network using the experimental event data from our electron pulse measurements. The network is composed of six fully connected layers with incrementally reducing layer sizes (200, 150, 100, 50, 25 and 10 knots per layer, ReLu activation function) down to a single regression output layer, as sketched in Fig. 3(a). For each event cluster, all ToA and ToT data, and event positions relative to the cluster center-of-mass are fed into the network. The number of input nodes is fixed corresponding to a maximum of 10 events per cluster. If less than 10 events per cluster are registered, the remaining input nodes are filled with default values. The scalar output of the network is the predicted arrival time of the incident electron relative to the trigger signal.

In the experiment, the arrival times of electrons at the detector relative to the photoemission laser pulses are fixed. To avoid training the network to this trivial case, for each event cluster a random number is added to the experimental ToA values and used as the respective ground truth. Overall, the network is trained with experimental data from $1.4 \times 10^5$ electron event cluster.

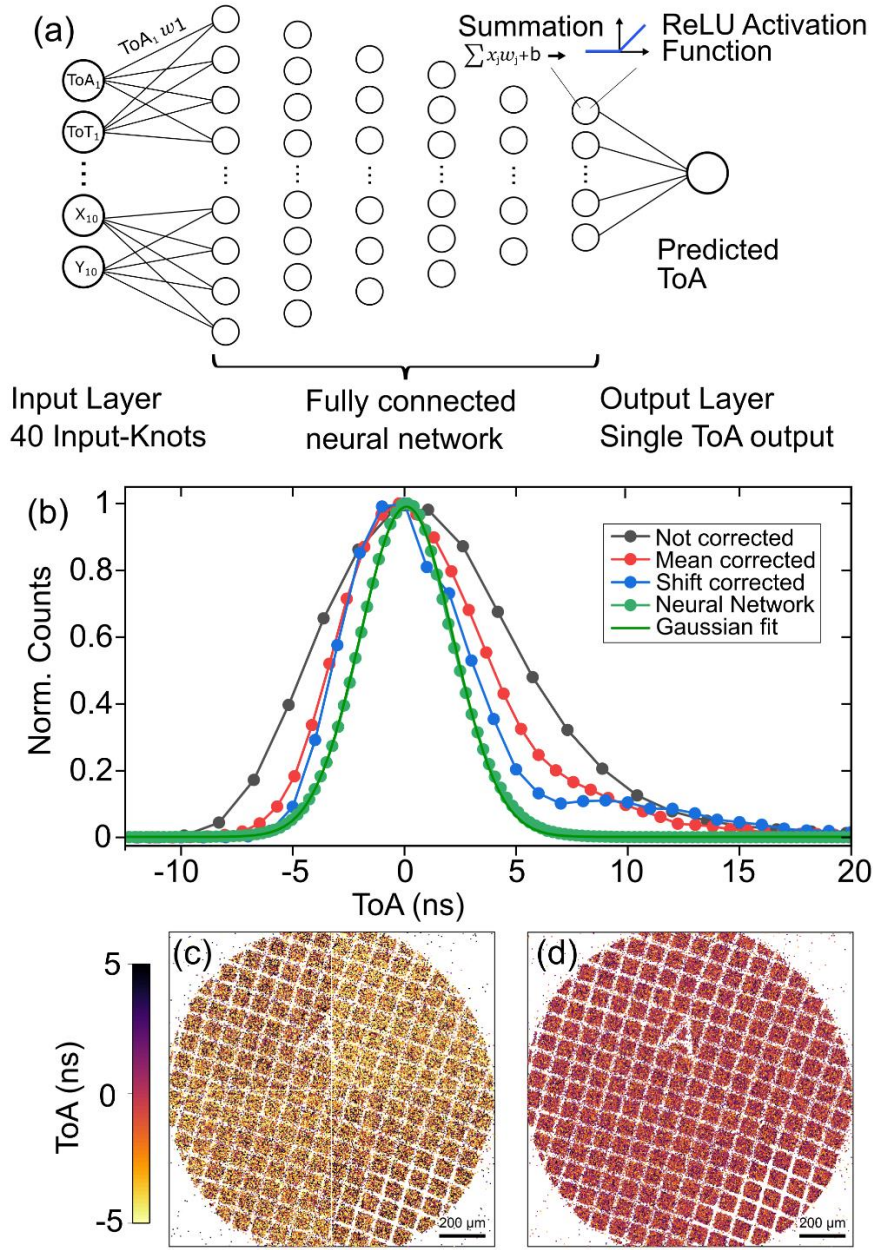

**Figure 3: Neural-network based timing prediction.** (a) Schematics of the neural network trained to predict the electron arrival time. (b) Histogram of the resulting time-of-arrival using uncorrected (gray), event cluster averaged (red), shift corrected (blue) and the neural network predicted data (green). See text for details. (c,d) Color-coded spatial distribution of arrival times using uncorrected (c) and neural network predicted (d) data. Color scale: time-of-arrival of an electron event.

In order to gauge the precision of the predicted arrival time, the neural network is introduced to experimental recordings not employed in the training process. Using recorded event data from $5 \times 10^6$ electron pulses, we obtained a ToA distribution of neural network corrected data, as shown in Fig. 3(b) (green curve). Notably, with our neural network approach, the predicted ToA distribution shows a root-mean-squared width of only 2 ns (4.7 ns FWHM) – a factor of 2 narrower compared to the uncorrected event data. Cluster averaging the ToA yields a 3.2 ns rms width, but also does not improve the tailing to larger time delays. The neural network corrects this effect, resulting in an approximately Gaussian shape of the distribution. We further note that including the absolute positions of each event in the neural network training data did not improve the accuracy of ToA prediction.

Finally, Fig. 3 (c,d) show a graphical representation of the achieved arrival time homogeneity across the detector for uncorrected pixel-based data (c) and neural network analyzed cluster data (d). In both cases, $2 \times 10^5$ detected events from the photoelectron beam were acquired and are represented by a dot at the position of the events with a color corresponding to the arrival time. The centered cross pattern in Fig. 3(c) is due to a 110-µm gap between the individual detector chips. As expected, applying the neural network results in a homogeneous timing information throughout the image (Fig 3 d). Additionally, by only considering the center of mass of each cluster, the gap between chips can be filled without noticeable artifacts.

## V. Conclusion

Using the precise timing of femtosecond photoelectron sources, we characterized correlations within the event-data stream of a TimePix3 electron detector. The event stream is used to train a neural network allowing to improve the precision of electron timing accuracies by a factor of two. Notably, this approach solely relies on experimental input data and requires no assumptions on the intricate scattering mechanism of fast electrons in the sensor layers or response behavior of the detector electronics. As such, we expect this approach to be equally applicable to other fast electron detectors with sub-nanosecond temporal resolution, currently being developed.

## Acknowledgments

We acknowledge financial support by the Volkswagen Foundation as part of the Lichtenberg Professorship "Ultrafast nanoscale dynamics probed by time-resolved electron imaging" and by the German Science Foundation within the grant INST 184/211-1 FUGG.

## Conflict of Interest

Two of the authors (LvV and MK) are paid employees of Amsterdam Scientific Instruments, the vendor of the TimePix3 electron detector characterized in this work.

## Data Availability

The data that support the findings of this study are available from the corresponding author upon reasonable request.

## References

[1] B. Barwick, H. S. Park, O.-H. Kwon, J. S. Baskin, and A. H. Zewail, "4D imaging of transient structures and morphologies in ultrafast electron microscopy," Science 322, 1227-1231 (2008).

[2] A. J. McKenna, J. K. Eliason, and D. J. Flannigan, "Spatiotemporal Evolution of Coherent Elastic Strain Waves in a Single MoS2 Flake," Nano letters 17, 3952-3958 (2017).

[3] Y.-J. Kim, Y. Lee, K. Kim, and O.-H. Kwon, "Light-Induced Anisotropic Morphological Dynamics of Black Phosphorus Membranes Visualized by Dark-Field Ultrafast Electron Microscopy," ACS nano 14, 11383-11393 (2020).

[4] C. Zhu, D. Zheng, H. Wang, M. Zhang, Z. Li, S. Sun, P. Xu, H. Tian, Z. Li, H. Yang, and J. Li, "Development of analytical ultrafast transmission electron microscopy based on laser-driven Schottky field emission," Ultramicroscopy 209, 112887 (2020).

[5] G. M. Vanacore, G. Berruto, I. Madan, E. Pomarico, P. Biagioni, R. J. Lamb, D. McGrouther, O. Reinhardt, I. Kaminer, B. Barwick, H. Larocque, V. Grillo, E. Karimi, F. J. García de Abajo, and F. Carbone, "Ultrafast generation and control of an electron vortex beam via chiral plasmonic near fields," Nature materials 18, 573-579 (2019).

[6] N. Rivera, G. Rosolen, J. D. Joannopoulos, I. Kaminer, and M. Soljačić, "Making two-photon processes dominate one-photon processes using mid-IR phonon polaritons," Proceedings of the National Academy of Sciences of the United States of America 114, 13607-13612 (2017).

[7] A. Feist, N. Bach, N. Da Rubiano Silva, T. Danz, M. Möller, K. E. Priebe, T. Domröse, J. G. Gatzmann, S. Rost, J. Schauss, S. Strauch, R. Bormann, M. Sivis, S. Schäfer, and C. Ropers, "Ultrafast transmission electron microscopy using a laser-driven field emitter: Femtosecond resolution with a high coherence electron beam," Ultramicroscopy 176, 63-73 (2017).

[8] K. Bücker, M. Picher, O. Crégut, T. LaGrange, B. W. Reed, S. T. Park, D. J. Masiel, and F. Banhart, "Electron beam dynamics in an ultrafast transmission electron microscope with Wehnelt electrode," Ultramicroscopy 171, 8-18 (2016).

[9] G. Cao, S. Jiang, J. _Akerman, and J. Weissenrieder, "Femtosecond laser driven precessing magnetic gratings," Nanoscale 13, 3746-3756 (2021).

[10] F. Houdellier, G. M. Caruso, S. Weber, M. Kociak, and A. Arbouet, "Development of a high brightness ultrafast Transmission Electron Microscope based on a laser-driven cold field emission source," Ultramicroscopy 186, 128-138 (2018).

[11] K. E. Priebe, C. Rathje, S. V. Yalunin, T. Hohage, A. Feist, S. Schäfer, and C. Ropers, "Attosecond electron pulse trains and quantum state reconstruction in ultrafast transmission electron microscopy," Nature Photonics 11, 793-797 (2017).

[12] D. Nabben, J. Kuttruff, L. Stolz, A. Ryabov, and P. Baum, "Attosecond electron microscopy of sub-cycle optical dynamics," Nature 619, 63–67 (2023).

[13] S. T. Park, M. Lin, and A. H. Zewail, "Photon-induced near-field electron microscopy (PINEM): theoretical and experimental," New Journal of Physics 12, 123028 (2010).

[14] A. Feist, K. E. Echternkamp, J. Schauss, S. V. Yalunin, S. Schäfer, and C. Ropers, "Quantum coherent optical phase modulation in an ultrafast transmission electron microscope", Nature 521, 200-203 (2015).

[15] R. Dahan, A. Gorlach, U. Haeusler, A. Karnieli, O. Eyal, P. Yousefi, M. Segev, A. Arie, G. Eisenstein, P. Hommelhoff, and I. Kaminer, "Imprinting the quantum statistics of photons on free electrons," Science 373, eabj7128 (2021).

16D. Jannis, K. Müller-Caspary, A. Béché, A. Oelsner, and J. Verbeeck, "Spectroscopic coincidence experiments in transmission electron microscopy," Applied Physics Letters 114, 143101 (2019).

[17] T. Weßels, S. Däster, Y. Murooka, B. Zingsem, V. Migunov, M. Kruth, S. Finizio, P.-H. Lu, A. Kovács, A. Oelsner, K. Müller-Caspary, Y. Acremann, and R. E. Dunin-Borkowski, "Continuous illumination picosecond imaging using a delay line detector in a transmission electron microscope," Ultramicroscopy 233, 113392 (2021).

[18] T. Poikela, J. Plosila, T. Westerlund, M. Campbell, M. de Gaspari, X. Llopart, V. Gromov, R. Kluit, M. van Beuzekom, F. Zappon, V. Zivkovic, C. Brezina, K. Desch, Y. Fu, and A. Kruth, "Timepix3: a 65K channel hybrid pixel readout chip with simultaneous ToA/ToT and sparse readout," Journal of Instrumentation 9, C05013-C05013 (2014).

[19] M. Ruat and C. Ponchut, "Characterization of a Pixelated CdTe X-Ray Detector Using the Timepix Photon-Counting Readout Chip," IEEE Trans. Nucl. Sci. 59, 2392-2401 (2012).

[20] D. Turecek, J. Jakubek, E. Trojanova, and L. Sefc, "Single layer Compton camera based on Timepix3 technology," Journal of Instrumentation 15, C01014-C01014 (2020).

[21] A. Zhao, M. van Beuzekom, B. Bouwens, D. Byelov, I. Chakaberia, C. Cheng, E. Maddox, A. Nomerotski, P. Svihra, J. Visser, V. Vrba, and T. Weinacht, "Coincidence velocity map imaging using Tpx3Cam, a time stamping optical camera with 1.5 ns timing resolution," The Review of scientific instruments 88, 113104 (2017).

[22] J. P. van Schayck, Y. Zhang, K. Knoops, P. J. Peters, and R. B. G. Ravelli, "Integration of an Event-driven Timepix3 Hybrid Pixel Detector into a Cryo-EM Workflow," Microscopy and Microanalysis 29, 352-363 (2023).

[23] J. P. van Schayck, E. van Genderen, E. Maddox, L. Roussel, H. Boulanger, E. Fröjdh, J.-P. Abrahams, P. J. Peters, and R. B. G. Ravelli, "Sub-pixel electron detection using a convolutional neural network," Ultramicroscopy 218, 113091 (2020).

[24] A. Feist, G. Huang, G. Arend, Y. Yang, J.-W. Henke, A. S. Raja, F. J. Kappert, R. N. Wang, H. Lourenço-Martins, Z. Qiu, J. Liu, O. Kfir, T. J. Kippenberg, and C. Ropers, "Cavity-mediated electron-photon pairs," Science 377, 777-780 (2022).

[25] H. Bromberger, C. Passow, D. Pennicard, R. Boll, J. Correa, L. He, M. Johny, C. C. Papadopoulou, A. Tul-Noor, J. Wiese, S. Trippel, B. Erk, and J. Küpper, "Shot-by-shot 250 kHz 3D ion and MHz photoelectron imaging using Timepix3," Journal of Physics B: Atomic, Molecular and Optical Physics 55, 144001 (2022).